\documentclass[runningheads]{llncs}
\usepackage[T1]{fontenc}
\usepackage{graphicx}

\usepackage{booktabs}
\usepackage{tabularx}
\usepackage{url}
\usepackage{hyperref}
\usepackage{amsmath}
\let\amsvec\vec
\let\vec\relax
\usepackage{MnSymbol}
\let\vec\amsvec

\usepackage{color}

\begin{document}
\title{MemoCare: An Interactive Multimodal Mobile System for Automated Cognitive Screening}
\titlerunning{MemoCare: Multimodal Mobile Cognitive Screening}
%

\author{
Duy-Cat Can\inst{1,2,3}
\and Mau Minh Phuc Le\inst{4}
\and Tuan-Khoa Hoang\inst{10}
\and Hai-Dang~Nguyen\inst{4}
\and Trung-Hieu Do\inst{5,6}
\and Dang Minh Ly\inst{7}
\and Minh-Duc Nguyen\inst{7,8}
\and Nghia TT Hoang\inst{7}
\and Nguyen Linh Trung\inst{3}$^\thinstar$
\and Huy-Hieu Pham\inst{4}$^\thinstar$
\and Huong Ha\inst{9,11}$^\thinstar$
\and Binh~T.~Nguyen\inst{10,11}$^\thinstar$
\and Oliver Y. Chén\inst{1,2}\thanks{Corresponding authors.}
}
\authorrunning{Can et al.}
%
\institute{
Lausanne University Hospital, Switzerland
\and Faculty of Biology and Medicine, University of Lausanne, Switzerland
\and VNU University of Engineering and Technology, Vietnam
\and VinUni-Illinois Smart Health Center, VinUniversity, Hanoi, Vietnam
\and Hanoi Medical University, Vietnam
\and National Geriatric Hospital, Vietnam
\and Department of Neurology, Military Hospital 175, Vietnam
\and Department of Neurology, School of Medicine, University of Medicine and Pharmacy at Ho Chi Minh City, Vietnam
\and International University, VNU-HCM, Vietnam
\and University of Science, VNU-HCM, Vietnam
\and Vietnam National University Ho Chi Minh City, Vietnam
\\
\email{
\{duy-cat.can,olivery.chen\}@chuv.ch,
htkhoa23@apcs.fitus.edu.vn,
\{hieu.ph,phuc.lmm,dang.nh3\}@vinuni.edu.vn,
dotrunghieu@hmu.edu.vn,
drminhdang@outlook.com,
nmduc@ump.edu.vn,
dr.hnghia@gmail.com,
linhtrung@vnu.edu.vn,
htthuong@hcmiu.edu.vn,
ngtbinh@hcmus.edu.vn
}
}
\maketitle              
\begin{abstract}
MemoCare is a React Native mobile system for automated multimodal cognitive screening in English and Vietnamese.
It integrates speech, temporal and spatial orientation, touchscreen interaction, and visuoconstruction.
Google Speech-to-Text transcripts are scored with deterministic, task-specific rules.
GPS-grounded current and historical administrative names support location scoring, touch events verify actions, and lightweight convolutional neural networks evaluate drawings.
A separate model provides visual interpretation.
Software verification passed 151/151 predefined language, spatial-answer, and touch-scoring regression cases and 48/48 fixed-coordinate spatial tests across four countries.
Transcript-based tests exclude speech recognition accuracy.
On a locked 71-image test set, validation-selected ShuffleNetV2 x1.5 achieved 91.33\% mean balanced accuracy and 78.87\% exact three-criterion accuracy.
Four clinician co-authors rated the workflow (pooled median 4/5 across eight criteria; item medians 3--4.5), highlighting improvements needed in speech handling, response acceptance, and result transparency.
The workflow automatically produces item-level and total scores.

\keywords{
    Cognitive screening
    \and Multimodal interaction
    \and Speech processing
    \and Geospatial reasoning
    \and Computer vision
}
\end{abstract}

\section{Introduction}
Cognitive screening brings different kinds of interaction together in a single assessment. Established tests such as the Mini-Mental State Examination (MMSE) include spoken responses, orientation to time and place, memory and calculation, commanded actions, language, and visuoconstruction~\cite{folstein1975mmse}. Mobile automation must therefore do more than display a questionnaire. It must capture different response types, interpret each according to its task-specific meaning, and combine them in a coherent scoring workflow.

Researchers have explored digital cognitive tests and speech-assisted mobile cognitive screening~\cite{chan2021diagnostic,devos2019mmse}, while multimodal research systems have combined sensing modalities to capture richer behavioral signals~\cite{jonell2021multimodal}. Deep learning has also been used to extract cognition-related features from copied drawings~\cite{tasaki2023pentamind}. MemoCare integrates task-specific scoring for speech, location, touch, and drawing in an interactive mobile workflow.
The contributions are:
(i) a unified mobile scoring architecture across four modalities;
(ii) task-specific processing modules;
and (iii) a bilingual prototype with technical verification and formative inspection by four clinician co-authors. Clinical diagnostic validity has not been evaluated.

\section{MemoCare System}

\begin{figure}[!t]
    \centering
    \includegraphics[width=\textwidth]{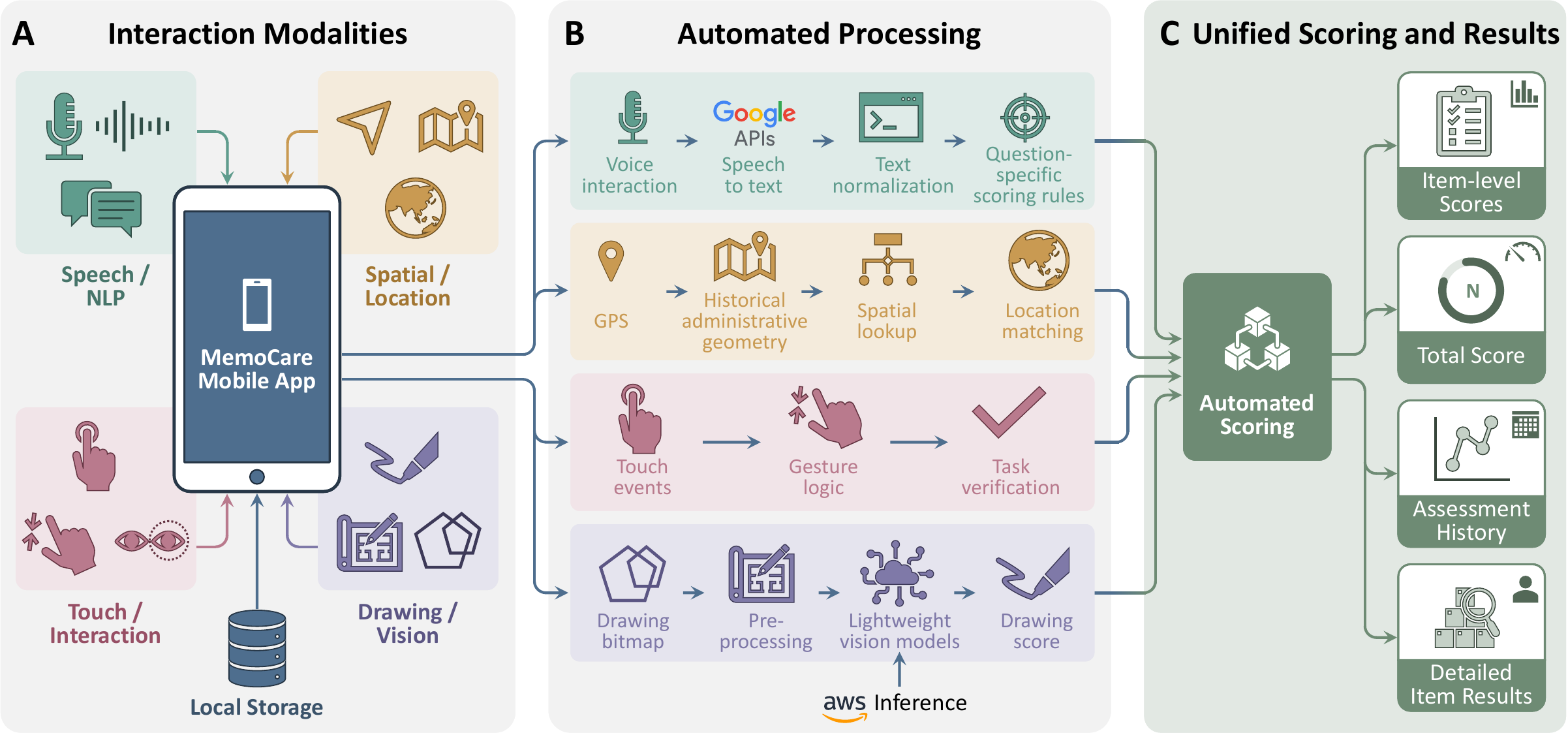}
    \caption{MemoCare architecture. (A) Speech, location, touch, and drawing are captured in one mobile app. (B) Specialized modules perform language scoring, geospatial reasoning, touch verification, and vision inference. (C) Item scores are aggregated into total score and local history.}
    \label{fig:architecture}
\end{figure}

MemoCare presents task-specific interactions and routes responses to processing modules (Fig.~\ref{fig:architecture}).
The workflow covers temporal and spatial orientation, memory, attention, calculation, language, touch-based commands, and visuoconstruction.
Some interactions are adapted for mobile input while preserving the cognitive operation.
Users can repeat, skip, pause, or resume tasks and view item-level and total scores afterward.

The React Native client supports Android and iOS.
Speech uses the microphone, touch and drawing use the display, and location uses device GPS.
Assessment state, scores, and history are stored locally.
Audio segments are sent to Google Speech-to-Text (STT) and drawing images to AWS-hosted inference, both processed by external providers.
A common item-scoring interface allows module updates without changing the mobile workflow.

The workflow runs without continuous examiner control.
Each task combines localized instructions, an input channel, and a completion condition.
If a response is missing, the app issues a reminder and allows the user to repeat or skip the task.
Stateful tasks retain earlier responses for sequential scoring.
English and Vietnamese use the same interaction flow, with language-specific scripts, normalization rules, and spatial vocabularies.

\section{Multimodal Processing and Scoring}

The assessment has four scoring paths: spoken responses for orientation, memory, calculation, and language; spoken place names matched to the device location; touch actions checked against event sequences or target regions; and geometric copying assessed from an image. Each pathway returns an item-level score.

\subsection{Speech and Language Scoring}
Most verbally answered tasks follow the same pipeline. Audio is trimmed and transcribed by Google STT, and the resulting transcript is normalized before task-specific scoring rules are applied. These rules cover date and number parsing, stateful sequential calculation, multi-target recall matching, ordered phrase matching, and sentence-completeness checks. They tolerate hesitations and repetitions, while language-specific normalization handles common transcription and lexical variants. For selected tasks, scoring prioritizes recognized task-relevant words and does not infer negation or correction from potentially incomplete STT transcripts.

The scoring module uses the normalized transcript together with the task state. This supports sequential responses while keeping speech recognition separate from task-specific scoring.

\subsection{Location-Aware Spatial Reasoning}
Spatial-orientation responses depend on the phone's position and may remain valid after administrative renaming or boundary changes. The MemoCare spatial module uses 89,043 administrative boundary records from Vietnam, Switzerland, China, and the United States. A spatial index retrieves candidate regions near the device location before point-in-polygon testing, avoiding a scan of the complete geometry collection.

For each resolved location, the module returns accepted current and historical place names. It normalizes these names and matches them against the STT transcript using exact and approximate matching, so administrative aliases and renamed regions can be accepted where appropriate. The administrative boundary data came from OpenStreetMap country extracts obtained through the Geofabrik download service.

\subsection{Touch Interaction}
For action tasks adapted to mobile input, MemoCare scores touchscreen events instead of speech. One interaction checks a multi-step tap-and-drag sequence by action type, target, and event order. Another checks whether a touch falls within a predefined target region on a displayed image. The module compares recorded events with task-specific interaction rules. All touch processing happens locally, without a camera or additional sensor.

\subsection{Drawing Representation and Data}
For the visuoconstruction task, the app displays a geometric reference for the user to reproduce on the touchscreen. Touch events could also be represented as stroke vectors, but the implemented scorer uses the rendered bitmap. This gives touch drawings and synthetic training images the same fixed image representation. The captured bitmap is resized to $224\times224$ pixels for convolutional neural network (CNN) inference. Automated analysis evaluates three internal structural criteria, C1--C3: component-shape validity and two spatial-relation conditions. The model uses these criteria for training and interpretation, while the application shows only the configured binary item result. The item is scored as positive only when all three criteria are satisfied.

Because the manually drawn training set was limited, we developed a rule-based generator to produce drawings with known C1/C2/C3 labels. The four generation stages produced 45,724 images covering valid and invalid structures, relation errors, hard negatives, and visual variations such as blur, perspective, shadow, texture, and missing edges.

The development dataset for this technical evaluation contained 329 training, 66 validation, and 71 locked-test drawings, labeled using the predefined C1/C2/C3 criteria. Because criterion distributions are imbalanced, we evaluated performance using balanced accuracy (BA), the average of sensitivity and specificity. Mean BA averages the three criterion-specific BAs. Exact three-criterion accuracy requires all C1--C3 predictions for a drawing to be correct.

\subsection{Vision Models and Deployment}
We evaluated eight lightweight CNN models with ImageNet-pretrained weights from four architecture families: ShuffleNetV2~\cite{ma2018shufflenetv2}, MobileNetV3~\cite{howard2019mobilenetv3}, EfficientNet-B0~\cite{tan2019efficientnet}, EfficientNet-Lite variants, and MnasNet~\cite{tan2019mnasnet}. Each model predicts C1--C3 probabilities, is first trained on the staged synthetic dataset, and is subsequently fine-tuned on the manually drawn training set.

For each model and criterion, we used the 66-image validation set to select probability calibration and decision thresholds. We compared raw and temperature-scaled probabilities~\cite{guo2017calibration} and used bootstrap resampling to identify stable threshold ranges before final evaluation.

The deployment consensus first evaluates MobileNetV3-Large and EfficientNet-B0. EfficientNet-Lite1 is invoked only when they disagree on at least one criterion, after which C1--C3 are determined by criterion-wise majority voting.
A stable AWS endpoint exposes the drawing inference service, keeping the mobile client independent of the selected vision backbone. A separate criterion-aware U-Net~\cite{ronneberger2015unet} provides a structured visual interpretation of component shapes and spatial-relation regions. This explanatory model does not affect the binary drawing score.

\section{Technical Verification and Expert Review}

\begin{table}[!t]
\caption{Technical verification and expert inspection.}
\label{tab:evaluation}
\centering
\small
\setlength{\tabcolsep}{3pt}
\begin{tabularx}{\textwidth}{@{}>{\raggedright\arraybackslash}p{0.23\textwidth} >{\raggedright\arraybackslash}X >{\raggedright\arraybackslash}p{0.33\textwidth}@{}}
\toprule
\textbf{Module} & \textbf{Evaluation} & \textbf{Result} \\
\midrule
Speech/language & Transcript-level scoring & 138/138 passed \\
Spatial & Fixed-coordinate and answer scoring & 48/48 and 8/8 passed \\
Touch & Interaction scoring & 5/5 passed \\
Drawing, single model & ShuffleNetV2 x1.5, locked test ($n=71$) & Mean BA 91.33\%; three-criterion exact 78.87\% \\
Drawing, consensus & Locked test ($n=71$), descriptive & Mean BA 95.17\% \\
Clinician inspection & Eight 5-point criteria, $n=4$
              & Median 4/5 (item medians 3--4.5) \\
\bottomrule
\end{tabularx}
\end{table}

\subsection{Technical Verification}
The developer-defined regression cases test whether a processing module returns the expected output for a specified input. A case passes when the computed score or resolved location matches its predefined expected result. The scoring suite passed all 151 cases: 138 speech/language, 8 spatial-answer, and 5 touch/action cases. Spatial regression also passed all 48 fixed-coordinate cases across the four supported countries. Because the speech/language cases start with STT transcripts, these results verify downstream scoring logic but do not measure end-to-end audio transcription or speech scoring accuracy. Table~\ref{tab:evaluation} summarizes the technical results and formative clinician inspection.

We used the 66-image validation set for drawing model selection, probability calibration, and threshold setting. ShuffleNetV2 x1.5 had the highest validation mean BA at 96.23\% and was selected as the single-model result. On the locked 71-image test set, it achieved 91.33\% mean BA and 78.87\% exact three-criterion accuracy. The locked test was not used for model selection, calibration, or threshold tuning.

After the model evaluation, we combined MobileNetV3-Large, EfficientNet-B0, and EfficientNet-Lite1 in a post-hoc consensus, based on their complementary performance across the completed evaluation results. The conditional consensus achieved a descriptive mean BA of 95.17\% on the locked test set. Since we configured the consensus after reviewing the locked-test results, this is a descriptive post-hoc estimate from those same 71 images.

\subsection{Formative Clinician Inspection}
Four clinician co-authors, one in geriatrics and three in neurology, each evaluated the complete application using an end-to-end workflow with mock responses.
They rated eight criteria on a 5-point scale: fidelity of the digital workflow, clarity of instructions, suitability of voice and touchscreen interaction, agreement of automated scoring with clinical judgment, transparency of item-level results, availability of repeat/skip/pause/resume controls, and readiness for formal clinical evaluation.

The pooled median rating was 4/5, and item-level medians ranged from 3 to 4.5.
Clinicians gave the highest ratings to the overall digital workflow, clarity of instructions, and touchscreen interaction, each with a median of 4.5/5.
Voice interaction and readiness for formal clinical evaluation each received a median of 4/5. Agreement of automated scoring with clinical judgment, availability of interaction controls, and transparency of item-level results received medians of 3.5, 3.5, and 3/5, respectively.
The separate overall-readiness rating had a median of 4/5 (range 3--4).
Reviewers mainly suggested improvements to speech recognition and response handling, acceptable-response handling and task instructions, drawing interaction, and access to recorded responses.
These descriptive results reflect the four clinicians' assessments of the implemented workflow. Patient-level diagnostic performance remains to be evaluated.

\section{Limitations and Conclusion}
The regression tests verify specified module behavior, but clinical scoring accuracy has not been measured. Transcript-based tests do not evaluate microphone capture or STT errors across different accents and acoustic conditions. Spatial resources cover four countries, and drawing results depend on synthetic training and a small held-out set. The post-hoc consensus needs independent evaluation. Cloud transcription and image inference send audio and drawing inputs to external services and require network access. Latency, offline alternatives, provider retention, and privacy safeguards need assessment before clinical use. No prospective patient assessment was conducted. Clinical testing requires appropriate consent and ethics procedures.

MemoCare brings speech, spatial context, touch, and drawing together in a mobile cognitive-screening workflow, using modality-specific processing and a common scoring interface. Technical verification and formative clinician inspection establish the prototype's functionality and identify priorities for clinical evaluation.


%
%
%
%
\bibliographystyle{splncs04}
\bibliography{refs}

\clearpage
\appendix

\section{Demonstration Video and Setup}
\label{app:demo}

\paragraph{Demonstration video.}
A demonstration video of up to 3 minutes is available at \url{https://youtu.be/INKW6qbg4io}. It shows the implemented mobile workflow, major interaction modalities, automated scoring, and the setup used for the conference demonstration.

\paragraph{What attendees will see.}
MemoCare runs on an Android phone or tablet. The live demo shows task delivery, spoken-response transcription and scoring, GPS-grounded location scoring, touchscreen task verification, conditional drawing consensus scoring with criterion-level visual interpretation, and item-level and total results. The complete workflow remains available, while a shortened path highlights the major modalities in approximately 3--5 minutes.

\paragraph{How attendees interact.}
Attendees operate the device themselves. They can answer a spoken task, provide a location-related response, perform a touchscreen interaction task, complete a drawing task, and inspect the generated scores. Tasks can be repeated or skipped, allowing each attendee to focus on selected modalities. The live-demo instructions will identify that speech segments and drawing images are processed using external services and request that attendees use mock answers without disclosing personal health information.

\paragraph{Setup requirements.}
The demo requires one Android phone or tablet, standard table space, a power outlet, and Wi-Fi. The device uses its built-in microphone, touchscreen, and GPS/location service. The conference network must permit Google STT service and the remote drawing inference. Screen mirroring to an external display is optional. No camera, additional sensors, dedicated audio equipment, VR/AR hardware, or special physical space is required.

\end{document}